\documentclass{article}
\usepackage{graphicx}
\usepackage{amsmath}

\input{tcilatex}

\begin{document}

\title{\textbf{New Heisenberg Relations in a }\\
\textbf{Non-commutative Geometry}}
\author{W. Chagas-Filho \\
Departamento de Fisica, Universidade Federal de Sergipe\\
SE, Brazil}
\maketitle

\begin{abstract}
We show how an induced invariance of the massless particle action can be
used to construct an extension of the Heisenberg canonical commutation
relations in a non-commutative space-time.
\end{abstract}

\section{Introduction}

\noindent Relativistic particle theory was usually viewed as a prototype
theory for string theory and general relativity. Modern developments however
indicate that particle theory may play a much more fundamental role in our
understanding of the physical world. Let us briefly see how this change of
perspective in relation to particle theory may emerge.

In the past years, there has been an intense research activity on the
connection between string theory and non-commutative geometry. It is
believed that in quantum theories containing gravity, space-time must change
its nature at distances comparable to the Planck scale. This is because
quantum gravity has an uncertainty principle which prevents one from
measuring positions to better accuracies than the Planck length [1]. These
effects would then be modeled by a non-vanishing commutation relation
between the space-time coordinates. Many developments in string and
superstring theories were reached in this direction [2,3,4]. Nowadays,
however, it became clear that supergravity and the five consistent
superstring theories known are just particular limits of a larger, and
partially known theory, which is usually called M-theory. The most promising
formulations of M-theory are based on the supersymmetric quantum mechanics
of a system of N Dirichlet 0-branes, also called D-particles, which are
considered to compose the membrane structure [5]. In these formulations, the
quantum string and gravitational dynamics are generated by collective
excitations of these N D-particles.

In these formulations of M-theory the non-commutativity of the space-time
coordinates arise as a consequence of a N$\times $N matrix regularization of
the membrane space-time coordinates. With this matrix regularization, the
formalism is expected to reproduce the area-preserving diffeomorphism
invariance of the membrane world-volume as an N$\rightarrow \infty $ limit
of SU(N). This, of course, has its dangers. For a closed membrane with a
spherical topology, SU(N) identifies two opposite points on the surface of
the sphere with a single point, whereas diffeomorphism invariance does not
[6]. It would therefore be desirable to have a formulation of M-theory in
which the non-commutativity of the space-time coordinates arises as a
natural consequence of the system's own dynamics. To try to investigate this
problem using D-brane technology may turn out to be a very awkward exercise.
In this work we give evidence that a considerable gain in comprehension of
this complex problem may be achieved by considering the much simpler case of
the massless relativistic particle.

An approach to the study of the relations between string dynamics and
particle dynamics, using the concept of space-time symmetry as an
investigation tool, was presented in [7]. In particular, it was explicitly
verified in [7] the existence of a very special string motion in the high
energy limit of the theory. In this special motion, each point of the string
moves as if it were a massless particle. Thorn [8] was the first to point
out that strings should be regarded as composite systems of more fundamental
point-like objects. More recently it was claimed [9] that the Nambu-Goto
string can be considered as a continuous limit of ordered discrete sets of
relativistic particles for which the tangential velocities are excluded from
the action.

The fact that string theory is contained in M-theory, complemented with the
existence of the above mentioned string motions, compel us to search for an
indication that non-commutative geometry should emerge from relativistic
particle dynamics. This is precisely the motivation of the present work.
Here we show how we may use the special-relativistic ortogonality condition
between the velocity and acceleration to induce a new invariance of the
massless particle action. This new invariance may be used to perform a
transition to new canonical commutation relations which depend on the choice
of an arbitrary function $\beta $ of the squared momentum. In particular,
the space-time positions now fail to commute. These new commutation
relations with arbitrary $\beta $ obey all Jacobi identities among the
canonical variables and reduce to the usual Heisenberg commutation relations
if the mass shell condition is imposed. After choosing a specific form for $%
\beta $ and computing the commutators, it becomes impossible to return to a
commutative geometry by imposing the mass shell condition. The main
conclusion of this work is that, for a massless relativistic particle, there
is a particular inertial frame in which the uncertainty introduced by two
simultaneous position measurements is equivalent to a space-time rotation.
This agrees with an earlier observation made in the context of string theory
[10].

\section{Relativistic Particles}

A relativistic particle describes is space-time a one-parameter trajectory $%
x^{\mu }(\tau )$. A possible form of the action is the one proportional to
the arc length traveled by the particle and given by 
\begin{equation}
S=-m\int ds=-m\int d\tau \sqrt{-\dot{x}^{2}}  \tag{2.1}
\end{equation}
In this work we choose $\tau $ to be the particle's proper time, $m$ is the
particle's mass and $ds^{2}=-\delta _{\mu \nu }dx^{\mu }dx^{\nu }$. A dot
denotes derivatives with respect to $\tau $ and we use units in which $\hbar
=c=1$.

This particle action is invariant under the reparametrization 
\begin{equation}
\tau \rightarrow \tau ^{\prime }=f(\tau )  \tag{2.2}
\end{equation}
where $f$ is an arbitrary continuous function of $\tau .$ Action (2.1)
defines the simplest generally covariant physical system.

Action (2.1) is obviously inadequate to study the massless limit of the
theory and so we must find an alternative action. Such an action can be
easily computed by treating the relativistic particle as a constrained
system. In the transition to the Hamiltonian formalism action (2.1) gives
the canonical momentum 
\begin{equation}
p_{\mu }=\frac{m}{\sqrt{-\dot{x}^{2}}}\dot{x}_{\mu }  \tag{2.3}
\end{equation}
and this momentum gives rise to the primary constraint 
\begin{equation}
\phi =\frac{1}{2}(p^{2}+m^{2})=0  \tag{2.4}
\end{equation}
which is the mass shell condition for the relativistic particle. We follow
Dirac's [11] convention that a constraint is set equal to zero only after
all calculations have been performed. The canonical Hamiltonian
corresponding to action (2.1), $H=p.\dot{x}-L$, identically vanishes. This
is a characteristic feature of generally covariant systems described with a
square root action. Dirac's Hamiltonian for the relativistic particle is
then 
\begin{equation}
H_{D}=H+\lambda \phi =\frac{1}{2}\lambda (p^{2}+m^{2})  \tag{2.5}
\end{equation}
where $\lambda (\tau )$ is a Lagrange multiplier. The Lagrangian that
corresponds to (2.5) is 
\begin{eqnarray}
L &=&p.\dot{x}-H_{D}  \notag \\
&=&p.\dot{x}-\frac{1}{2}\lambda (p^{2}+m^{2})  \TCItag{2.6}
\end{eqnarray}
Solving the equation of motion for $\ p_{\mu }$ that \ follows from (2.6)
and inserting the result back in it, we obtain the particle action 
\begin{equation}
S=\int d\tau (\frac{1}{2}\lambda ^{-1}\dot{x}^{2}-\frac{1}{2}\lambda m^{2}) 
\tag{2.7}
\end{equation}
Now, if we solve the classical equation of motion for $\lambda (\tau )$ that
follows from (2.7)\ and insert the result back in it, we recover the
original action (2.1). Actions (2.7) and (2.1) are therefore classically
equivalent. The great advantage of (2.7) is that it has a smooth transition
to the $m=0$ limit.

Varying $x^{\mu }$ in (2.7) we obtain the classical equation for free motion 
\begin{equation}
\frac{d}{d\tau }(\lambda ^{-1}\dot{x}^{\mu })=\frac{dp^{\mu }}{d\tau }=0 
\tag{2.8}
\end{equation}

Now we make a transition to the massless limit. This limit is described by
the action 
\begin{equation}
S=\frac{1}{2}\int d\tau \lambda ^{-1}\dot{x}^{2}  \tag{2.9}
\end{equation}
The massless limit is usually considered to describe the high energy limit
of particle theory. The canonical momentum conjugate to $x^{\mu }$ is 
\begin{equation}
p_{\mu }=\lambda ^{-1}\dot{x}_{\mu }  \tag{2.10}
\end{equation}
and will lead to the same classical equation of motion (2.8) satisfied by
the particle with a non-vanishing mass. The canonical momentum conjugate to $%
\lambda $ identically vanishes and this is a primary constraint, $p_{\lambda
}=0$. Constructing the canonical Hamiltonian, and requiring the stability of
this constraint, we obtain the secondary constraint 
\begin{equation}
\phi =\frac{1}{2}p^{2}=0  \tag{2.11}
\end{equation}
which is the mass shell condition for the massless particle.

Now we may use the fact that we are dealing with a special-relativistic
system. Special relativity has the kinematical feature that the relativistic
velocity is always ortogonal to the relativistic acceleration [12]. We can
use this ortogonality to induce the invariance of the massless particle
action (2.9) under the transformation 
\begin{equation}
x^{\mu }\rightarrow \tilde{x}^{\mu }=\exp \{\beta (\dot{x}^{2})\}x^{\mu } 
\tag{2.12a}
\end{equation}
\begin{equation}
\lambda \rightarrow \exp \{2\beta (\dot{x}^{2})\}\lambda  \tag{2.12b}
\end{equation}
where $\beta $ is an arbitrary function of $\dot{x}^{2}$. We emphasize that
although the ortogonality condition must be used to get the invariance of
action (2.9) under transformation (2.12), this condition is not an external
ingredient in the theory. In fact, the ortogonality between the relativistic
velocity and acceleration is an unavoidable condition here. It is an
imposition of special relativity.

It may be pointed out that since the relativistic ortogonality condition is
used to turn the massless action invariant under (2.12), these
transformations may not be a true invariance of the action, being at most
only a symmetry of the equation of motion (2.8). It can be verified that
while the classical equation of motion (2.8), which is valid for both the
massive and massless particles, is invariant under transformation (2.12),
the massive particle action (2.7) is not. This reveals that transformation
(2.12) is a symmetry of the equations of motion only in the massive theory.
In the massless theory it is a symmetry of the equations of motion and of
the action.

Now we consider the commutator structure that transformation (2.12a)\
induces in the massless particle's phase space. We begin by assuming the
usual Heisenberg commutation relations between the canonical variables, $%
[x_{\mu },x_{\nu }]=[p_{\mu },p_{\nu }]=0$ , $[x_{\mu },p_{\nu }]=i\delta
_{\mu \nu }$. Taking $\beta (\dot{x}^{2})=\beta (\lambda ^{2}p^{2})$ in
transformation (2.12a) and transforming the $p_{\mu }$ in the same manner as
the $x_{\mu },$ we find that the new transformed canonical variables $(%
\tilde{x}_{\mu },\tilde{p}_{\mu })$ obey the commutators 
\begin{equation}
\lbrack \tilde{p}_{\mu },\tilde{p}_{\nu }]=0  \tag{2.13}
\end{equation}
\begin{equation}
\lbrack \tilde{x}_{\mu },\tilde{p}_{\nu }]=i\delta _{\mu \nu }(1+\beta
)^{2}+(1+\beta )[x_{\mu },\beta ]p_{\nu }  \tag{2.14}
\end{equation}
\begin{equation}
\lbrack \tilde{x}_{\mu },\tilde{x}_{\nu }]=(1+\beta )\{x_{\mu }[\beta
,x_{\nu }]-x_{\nu }[\beta ,x_{\mu }]\}  \tag{2.15}
\end{equation}
written in terms of the old canonical variables. These commutators obey the
non trivial Jacobi identities $(\tilde{x}_{\mu },\tilde{x}_{\nu },\tilde{x}%
_{\lambda })=0$ \ and \ $(\tilde{x}_{\mu },\tilde{x}_{\nu },\tilde{p}%
_{\lambda })=0$. They also reduce to the usual Heisenberg commutators when $%
\beta (\lambda ^{2}p^{2})=0$.

It may be questioned here, based on the definition (2.10) of the classical
momentum, why the quantum momentum should transform in the same way as $%
x^{\mu }$ under transformation (2.12a). The identification of the correct
quantum canonical momentum is a typical problem in non-commutative field
theories [13]. An example is non-commutative quantum string field theory,
where the momentum space Fourier transform of the odd vibrational position
modes of the string behave as the quantum canonical momenta conjugate to the
even vibrational position modes [3]. In the context of this work,
transforming the quantum momenta in the opposite way of the $x^{\mu }$, as
is suggested by the definition (2.10) of the classical momentum, implies
substituting the commutator (2.14) by 
\begin{equation*}
\lbrack \tilde{x}_{\mu },\tilde{p}_{\nu }]=(1+\beta )\{i\delta _{\mu \nu
}(1-\beta )-[x_{\mu },\beta ]p_{\nu }\}
\end{equation*}
This commutator also reduces to the usual Heisenberg commutator when the
mass shell condition $\beta =0$ is imposed. But the Jacobi identity $(\tilde{%
x}_{\mu },\tilde{x}_{\nu },\tilde{p}_{\lambda })=0$ based on commutator
(2.15) and the above commutator does not seem to close. If this is correct,
the classical and quantum canonical momenta of the massless particle on a
non-commutative space-time are related by a transformation of the type
(2.12). That is $p_{q}^{\mu }=\exp \{2\beta \}p_{c}^{\mu }$.

The simplest example of a non-commutative geometry induced by transformation
(2.12) is the case when $\beta (\lambda ^{2}p^{2})=\lambda ^{2}p^{2}$.
Computing the commutators (2.14) and (2.15) for this form of $\beta ,$ and
imposing the mass shell condition (2.11), we arrive at the new canonical
commutation relations 
\begin{equation}
\lbrack \tilde{p}_{\mu },\tilde{p}_{\nu }]=0  \tag{2.16}
\end{equation}
\begin{equation}
\lbrack \tilde{x}_{\mu },\tilde{p}_{\nu }]=i\delta _{\mu \nu }+i\lambda
^{2}p_{\mu }p_{\nu }  \tag{2.17}
\end{equation}
\begin{equation}
\lbrack \tilde{x}_{\mu },\tilde{x}_{\nu }]=-2i\lambda ^{2}M_{\mu \nu } 
\tag{2.18}
\end{equation}
where $M_{\mu \nu }=x_{\mu }p_{\nu }-x_{\nu }p_{\mu }$ is the generator of
Lorentz rotations. The commutator (2.18) satisfies the condition 
\begin{equation}
\int \mathbf{Tr}[\tilde{x}_{\mu },\tilde{x}_{\nu }]=0  \tag{2.19}
\end{equation}
as is the case for a general non-commutative algebra [14]. From equation
(2.18) we see that there is an inertial frame in which the uncertainty
introduced by two simultaneous position measurements is equivalent to a
rotation in space-time. It is clear from (2.17) and (2.18) that we recover
the usual Heisenberg canonical commutator relations only when $\lambda (\tau
)$ vanishes.

\bigskip

\textbf{Acknowledgment-} The author thanks Dr. R. P. Malik, Dr. S. Ghosh,
Dr. M. S. Plyushchay and Dr. P. Mukherjee for their nice influence in the
development of this work.

\end{document}